\documentclass[aps,prb,preprint,groupedaddress,showpacs]{revtex4}

\usepackage{amsmath,amssymb,latexsym,epic,eepic,epsfig,graphics,psfrag}
\usepackage[latin1]{inputenc}
\newcommand{\eq}[1]{Eq.~(\ref{#1})}
\newcommand{\fig}[1]{Fig.~\ref{#1}}
\newcommand{\tab}[1]{Table~\ref{#1}}
\newcommand{\eqs}[1]{Eqs.~(\ref{#1})}

\newcommand{\sect}[1]{Sect.~\ref{#1}}

\newcommand{\rbt}{\mathring{\mathbf{t}}}
\newcommand{\tbt}{\tilde{\mathbf{t}}}
\newcommand{\ba}{\mathbf{a}}

\newcommand{\br}{\mathbf{r}}
\newcommand{\bt}{\mathbf{t}}
\newcommand{\bb}{\mathbf{b}}

\newcommand{\bB}{\mathbf{B}}

\newcommand{\bX}{\mathbf{X}}

\newcommand{\bH}{\mathbf{H}}

\newcommand{\bPhi}{\mathbf{\Phi}}
\newcommand{\tbPhi}{\tilde{\mathbf{\Phi}}}
\newcommand{\bS}{\mathbf{S}}
\newcommand{\bI}{\mathbf{I}}

\newcommand{\bLambda}{\mathbf{\Lambda}}

\newcommand{\tbLambda}{\tilde{\mathbf{\Lambda}}}
\newcommand{\tba}{\tilde{\mathbf{a}}}

\newcommand{\mt}{\mathrm{t}}

\newcommand{\mk}{\mathrm{k}}

\newcommand{\mK}{\mathrm{K}}

\newcommand{\mT}{\mathrm{T}}
\newcommand{\mG}{\mathrm{G}}

\usepackage{revsymb}
\usepackage{bm}
\usepackage{comment}

\begin{document}

\title{Efficient wave function matching approach for quantum transport calculations}

\author{Hans Henrik B. S\o{}rensen}
 \email{hhs@imm.dtu.dk}
\affiliation{Informatics and Mathematical Modelling,
  Technical University of Denmark, 
  Bldg. 321, DK-2800 Lyngby, 
  Denmark}
\author{Dan Erik Petersen}
\affiliation{Department of Computer Science, 
  University of Copenhagen,
  Universitetsparken 1, 
  DK-2100 Copenhagen, Denmark}
\author{Per Christian Hansen}
\affiliation{Informatics and Mathematical Modelling,
  Technical University of Denmark, 
  Bldg. 321, DK-2800 Lyngby, 
  Denmark}
\author{Stig Skelboe}
\affiliation{Department of Computer Science, 
  University of Copenhagen,
  Universitetsparken 1, 
  DK-2100 Copenhagen, 
  Denmark}
\author{Kurt Stokbro}
\affiliation{Department of Computer Science, 
  University of Copenhagen,
  Universitetsparken 1, 
  DK-2100 Copenhagen, Denmark}

\date{\today}

\begin{abstract}
The Wave Function Matching (WFM) technique has recently been developed
for the calculation of electronic transport in quantum two-probe
systems. In terms of efficiency it is comparable with the widely used 
Green's function approach.
The WFM formalism presented so far requires the evaluation of
all the propagating and evanescent bulk modes of the left and right
electrodes in order to obtain the correct
coupling between device and electrode regions.
In this paper we will describe a modified WFM approach that
allows for the exclusion of the vast majority of
the evanescent modes in all parts of the calculation.
This approach makes it feasible to apply iterative techniques 
to efficiently determine the few required bulk modes, which
allows for a significant reduction of the computational expense of the
WFM method. We illustrate the efficiency of the method on
a carbon nanotube field-effect-transistor (FET) device
displaying band-to-band tunneling and modeled
within the semi-empirical Extended Hückel 
theory (EHT) framework.
\end{abstract}

\maketitle

\section{\label{sec:theory}Introduction}

Quantum transport simulations have become an important theoretical
tool for investigating the electrical properties of nano-scale
systems. \cite{Datta2005,Brandbyge2002,Buttiker1985,Meir1992,Reed1997} 
The basis for the approach is the Landauer-Büttiker picture
of coherent transport, where the electrical properties of a nano-scale
constriction is described by the transmission coefficients of a
number of one-electron modes propagating coherently through the
constriction. The approach has been used successfully to describe the
electrical properties of a wide range of nano-scale systems, including
atomic wires, molecules and
interfaces. \cite{Faleev2005,Pomorski2004,Gokturk2005,Stilling2006,Nitzan2003,DiVentra2000,Stokbro2003,Lang2000,Larade2001,Khomyakov2004}
In order to apply the method
to semiconductor device simulation, it is necessary to handle systems
comprising many thousand atoms, and this will require new efficient
algorithms for calculating the transmission coefficient. 

Our main purpose in this paper is to give details of a
method we have developed, based on the
WFM technique, \cite{Ando1991,Khomyakov2005,Brocks2007}
which is suitable for studying
electronic transport in large-scale atomic
two-probe systems, such as large carbon nanotubes or nano-wire configurations.

\begin{figure}[tbp]
\begin{center}
\includegraphics*[width=8.3cm]{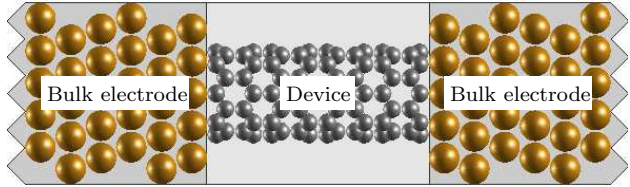}
\end{center}
\caption{(Color online)
Schematic illustration of a nano-scale two-probe system in which
a device is sandwiched between two semi-infinite 
bulk electrodes.
}
\label{fig:twoprobe}
\end{figure}

We adopt the many-channel formulation of Landauer and B\"uttiker
to describe electron transport in nano-scale two-probe systems
composed of a left and a right electrode attached 
to a central device, see \fig{fig:twoprobe}.
In this formulation, the conduction ${\cal G}$ of 
incident electrons through the device is
intuitively given in terms of transmission and reflection matrices, 
$\bt$ and $\br$, that satisfy the unitarity condition
$\bt^\dagger \bt + \br^\dagger \br = \bm{1}$
in the case of elastic scattering.
The matrix element $t_{ij}$ is the probability amplitude of an
incident electron in a mode $i$ in the left electrode being
scattered into a mode $j$ in the right electrode, and correspondingly
$r_{ik}$ is the probability of it
being reflected back into mode $k$ in the left electrode.
This simple interpretation yields the Landauer-B\"uttiker 
formula \cite{Buttiker1985}
\begin{equation}\label{eqn:landauer}
{\cal G} = \frac{2 e^2 }{h} \mathrm{Tr}[ \bt^\dagger \bt ],
\end{equation}
which holds in the limit of infinitesimal voltage bias and zero temperature.

To our knowledge, the 
WFM schemes presented so far in the literature require 
the evaluation of all the Bloch and evanescent bulk modes of 
the left and right electrodes in order to obtain the correct
coupling between device and electrode regions. The reason for this
is that the complete set of bulk modes is needed to
be able to represent the proper reflected and transmitted wave functions.
In this paper we will describe a modified WFM approach that
allows for the exclusion of the vast majority of
the evanescent modes in all parts of the calculation.
The primary modification can be pictured as a simple
extension of the central region with a few principal electrode
layers. In this manner, it becomes advantageous to apply
iterative techniques for obtaining the relatively few
Bloch modes and slowly decaying evanescent modes that are required. 
We have recently developed such an iterative method in 
Ref.~\onlinecite{Sorensen2008}, which 
allows for an order of magnitude reduction of the computational expense of the
WFM method in practice.

In this work, the proper analysis of the modified WFM approach is
presented. The accuracy of the method is investigated and
appropriate error estimates are developed.
As an illustration of the applicability of our WFM scheme we
consider a 1440 atom CNTFET device of 14 nm in length. We 
calculate the zero-bias transmission curves of the device under 
various gate voltages and reproduce previously established 
characteristics of band-to-band tunneling.\cite{Appenzeller2004} 
We compare directly the results of the modified WFM method to those of
the standard WFM method for quantitative verification of the
calculations.  

The rest of the paper is organized as follows.
The WFM formalism used to obtain 
$\bt$ and $\br$ is
introduced in Sect.~\ref{sec:formalism}.
In Sect.~\ref{sec:excluding} we 
present our method to effectively exclude the rapidly decaying
evanescent modes from the two-probe transport
calculations.
Numerical results are
presented in Sect.~\ref{sec:numerical}.
and the paper ends with a short summary and
outlook.

\section{\label{sec:formalism}Formalism}

In this section we give a minimal review
of the formalism and notation that is used in
the current work in order to determine the transmission and reflection matrices 
$\bt$ and $\br$.
This WFM technique has several attractive features compared to the
widely used and mathematically equivalent Green's function approach.
\cite{Datta2005, Brandbyge2002} Most importantly,
the transparent Landauer picture of electrons scattering via the central region
between Bloch modes of the electrodes is retained throughout the
calculation. Moreover, WFM allows one to consider the
significance of each available mode
individually in order to achieve more efficient numerical 
procedures to obtain $\bt$ and $\br$.

\subsection{\label{sec:wfm}Wave function matching}

The WFM method is based upon direct matching
of the bulk modes in the left and right electrode to the
scattering wave function of the central region. For the most part
this involves two major tasks; obtaining the bulk electrode modes and
solving a system of linear equations. 
The bulk electrode modes can be 
characterized as either propagating or
evanescent (exponentially decaying) modes but 
only the propagating modes contribute to
${\cal G}$ in \eq{eqn:landauer}. We may write
${\cal G} = (2 e^2/h) T$, where 
\begin{equation}\label{eqn:T}
T = \sum_{kk^\prime} |t_{kk^\prime}|^2
\end{equation}
is the total transmission and
the sum is limited to propagating modes $k$ and $k^\prime$ in the
left and right electrode, respectively. Notice, however,
that the evanescent modes are still needed in order to obtain the correct
matrix elements $t_{kk^\prime}$. We will discuss this matter in
\sect{sec:accuracy}.

\begin{figure}[tbp]
\begin{center}
\includegraphics*[width=8.3cm]{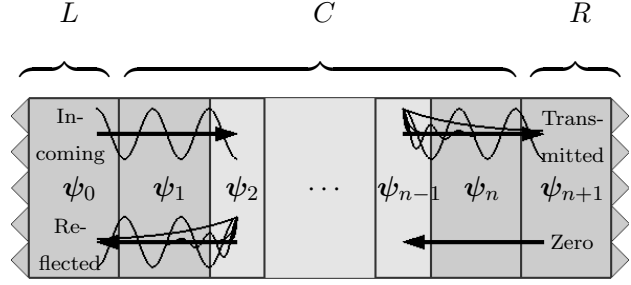}
\end{center}
\caption{(Color online)
Schematic representation of WFM applied to layered two-probe systems,
where the central device region, consisting of layers $i=1,\dots,n$,
is attached to left and right semi-infinite electrodes.
The incoming propagating mode from the left electrode
is scattered in the central region and ends up as
reflected and transmitted 
superpositions of propagating and evanescent modes.
}
\label{fig:wfm}
\end{figure}

We assume a tight-binding setup for the two-probe systems
in which the infinite structure is divided into principal layers 
numbered $i = -\infty,\dots,\infty$ and composed of a
finite central ($C$) region containing the device and two semi-infinite
left ($L$) and right ($R$) electrode regions, see \fig{fig:wfm}.
The wave function is
$\bm{\psi}_i(\bm{x}) = \sum_j^{m_i} c_{i,j}
\chi_{i,j}(\bm{x}-\bX_{i,j})$ in layer $i$,
where $\chi_{i,j}$ denotes localized non-orthogonal atomic 
orbitals and $\bX_{i,j}$ are the positions of the $m_i$ orbitals in layer $i$.
We represent $\bm{\psi}_i(\bm{x})$
by a column vector of the expansion coefficients, given by
$\bm{\psi}_i=[c_{i,1}, \dots, c_{i,m_i}]^\mathrm{T}$,
and write the wave function $\bm{\psi}$ extending over the entire
system as $\bm{\psi} = [\bm{\psi}_{-\infty}^\mathrm{T},
\dots, \bm{\psi}_{\infty}^\mathrm{T}]^\mathrm{T}$. 
We also assume that the border layers $1$ and $n$ of the central
region are always identical to a layer of the connecting electrodes.

We refer the reader to
Refs.~\onlinecite{Ando1991,Khomyakov2005,Brocks2007,hhsphd2008} 
for details on how to
employ WFM to our setup.
Here and in the rest of this paper, we will 
use the following notation for the key elements:
The matrices
$\bm{\Phi}_{L}^\pm = [\bm{\phi}_{L,1}^{\pm},\dots,\bm{\phi}_{L,m_L}^{\pm}]$
contain in their columns the full set of $m_L$ left-going ($-$) 
and $m_L$ right-going ($+$) bulk modes 
$\bm{\phi}_{L,k}^{\pm}$ of the left electrode, and the diagonal matrices
$\bm{\Lambda}_L^\pm = \mathrm{diag}[
\lambda_{L,1}^{\pm}, \lambda_{L,2}^{\pm}, \dots, \lambda_{L,m_L}^{\pm}]$
hold the corresponding Bloch factors.
\footnote{
Bloch's theorem\cite{Ashcroft1976}
$\bm{\psi}_i = \lambda_k \bm{\psi}_{i-1}$ for the ideal electrodes
defines the phase factors
$\lambda_k \equiv e^{\imath q_kd}$, 
where $q_k$ is the complex wave number and $d$ is the layer thickness,
which are referred to as Bloch factors throughout this paper.
}
If trivial modes with $|\bm{\phi}_{L,k}^+|=\bm{0}$ or
$|\bm{\phi}_{L,k}^-|=\tt{\bm{\infty}}$
occur they are simply rejected.
We assume that all the evanescent bulk modes are (state-)normalized
$\bm{\phi}_{L,k}^{\pm\dagger}\bm{\phi}^\pm_{L,k} = 1$, while all the
Bloch bulk modes are
flux-normalized
\footnote{
When using the Landauer 
formula in \eq{eqn:landauer} it is assumed that the
electrode Bloch modes carry
unit current in the conduction direction. This can be
conveniently accommodated by flux-normalizing the 
Bloch modes, i.e., 
$\bm{\phi}_{L,k}^\pm
\rightarrow (d_{L}/v_{L,k}^\pm)^\frac{1}{2} \bm{\phi}_{L,k}^\pm$,
in the case of the left electrode.\cite{Fisher1981}}
$\bm{\phi}_{L,k}^{\pm\dagger}\bm{\phi}^\pm_{L,k} = d_{L}/v_{L,k}^\pm$,
where $v_{L,k}^\pm$ are the group velocities\cite{Ashcroft1976,Khomyakov2004}
and $d_L$ is the layer thickness. 
Similarly for the right
electrode the matrices $\bm{\Phi}^{\pm}_{R}$ and
$\bm{\Lambda}^{\pm}_{R}$ are formed.

We also introduce  the Bloch matrices\cite{Khomyakov2005}
$\bB_{L}^\pm = \bm{\Phi}_{L}^{\pm} 
\bm{\Lambda}_{L}^{\pm} (\bm{\Phi}_{L}^{\pm})^{-1}$
and
$\bB_{R}^\pm = \bm{\Phi}_{R}^{\pm} 
\bm{\Lambda}_{R}^{\pm} (\bm{\Phi}_{R}^{\pm})^{-1}$. 
which  propagate the layer wave functions 
in the bulk electrode
\begin{equation}\label{eqn:prop}
\bm{\psi}_{j}^\pm = (\bB^\pm)^{j-i} \bm{\psi}_{i}^\pm,
\end{equation}
where subscript $L$ is implied for the left electrode ($i,j \le 1$), 
and $R$ for the right electrode ($i,j \ge n$).  Notice that the first
central region layer is defined for layer $1$
and not layer $0$, as is the case in Ref.~\onlinecite{Brocks2007}. 

As explicitly shown in Refs.~\onlinecite{Ando1991,Khomyakov2005,Brocks2007},
by fixing the layer wave functions coming into the
$C$ region (e.g., in our case 
$\bm{\psi}_1^{+} = \lambda_{L,k}^{+} \bm{\phi}_{L,k}^{+}$
and $\bm{\psi}_n^{-} = \bm{0}$) and
matching the layer wave functions across the $C$ region
boundaries, the   system of linear equations for the central region wavefunction $\bm{\psi}_C$
can be  written as
\begin{equation}\label{eqn:linear}
(E\bS_C-\bH_C-\bm{\Sigma}_L-\bm{\Sigma}_R)\bm{\psi}_C=\bb,
\end{equation}
where $E$ is the energy, $\bS_C$ the overlap  and
$\bH_{C}$ the Hamiltonian matrix of the central
region. In the following we discuss the terms, $\bm{\Sigma}_L$,
$\bm{\Sigma}_R$, and $\bb$, which arise from matching the boundary
conditions with the electrode modes. 

The self-energy  matrices, $\bm{\Sigma}_L$ and $\bm{\Sigma}_R$,
arise from matching with the outgoing left and right electrode
modes. They only have non-zero terms in the upper left and lower right corner
block, respectively, and these elements can be calculated in terms of
the Bloch matrices:\cite{Ando1991,Khomyakov2005}
\begin{equation}\label{eqn:blochL}
[\bm{\Sigma}_L]_{1,1}= \bar{\bH}_{0,1}^\dagger 
(\bar{\bH}_{1}+\bar{\bH}_{0,1}^\dagger
(\bB_L^-)^{-1} )^{-1}
\bar{\bH}_{0,1},
\end{equation}
and
\begin{equation}\label{eqn:blochR}
[\bm{\Sigma}_R]_{n,n}= \bar{\bH}_{n,n+1} 
(\bar{\bH}_{n}+\bar{\bH}_{n,n+1}
\bB_R^+)^{-1}
\bar{\bH}_{n,n+1}^\dagger,
\end{equation}
where we have introduced the overline notation $\bar{\bH}_i \equiv E\bS_i
-\bH_i$ and $\bar{\bH}_{i,j} \equiv E\bS_{i,j}
-\bH_{i,j}$. For the  current setup, these matrices are 
\textit{identical} to the 
self-energy matrices introduced in the Green's function formalism
\cite{Datta2005} (to within an infinitesimal imaginary shift of $E$),
and may be evaluated by well-known
recursive techniques\cite{Guinea1983,Sancho1985} or constructed directly from the electrode modes using \eq{eqn:blochR}.

The source term $\bb$ arises from the incoming mode. Assuming an incoming mode from the left, we have 
$\bb = [\bb_{1}^\mathrm{T}, \bm{0}^\mathrm{T},
\dots, \bm{0}^\mathrm{T}]^\mathrm{T}$ specified by the expression
\begin{equation}\label{eqn:b1}
\bb_1 = -(\bar{\bH}_{0,1}^\dagger + [\bm{\Sigma}_L]_{1,1}
\bB_L^+ )\bm{\psi}_0,
\end{equation}
where $\bm{\psi}_0$ is the incoming wave function.

For notational simplicity in the following sections, we leave out the implied
subscripts $L$ or $R$, indicating the left 
or right electrode, whenever the formalism is the same for
both (e.g, for symbols $m, \lambda_k, \bm{\phi}_k,\bm{\Phi}^\pm,\bm{\Lambda}^\pm,\bB^\pm, \bm{\Sigma}$, etc.).

\subsection{\label{sec:tandr}Transmission and reflection coefficients}

As a final step we want to determine
the $\bt$ and $\br$ matrices from
the boundary wave functions $\bm{\psi}_1$
and $\bm{\psi}_n$ that have been obtained by solving \eq{eqn:linear}. 

When the incoming wave $\bm{\psi}_0$ is specified to be
the $k$th right-going mode $\bm{\phi}_{L,k}^{+}$ 
of the left electrode, then $\bm{\psi}_n$ will be the superposition of outgoing right transmitted waves. The  
 $k$th column of the transmission matrix $\bt_k$ is defined as  the corresponding expansion coefficients in right electrode modes and can be evaluated by solving
\begin{equation}\label{eqn:t}
\bm{\Phi}_R^+ \bt_k= \bm{\psi}_n,
\end{equation}
where $\bm{\Phi}_R^+$ is the $m_R \times m_R$ column matrix holding the right-going
bulk modes of the right electrode (and here assumed to be non-singular).
Similarly the $k$th column of the reflection matrix $\br_k$ is given by
\begin{equation}\label{eqn:r}
\bm{\Phi}_L^- \br_k= \bm{\psi}_1-\lambda_{L,k}^{+} 
\bm{\phi}_{L,k}^{+},
\end{equation}
where $\bm{\Phi}_L^-$ holds the left-going bulk modes of the left electrode.
The flux normalization ensures that 
$\bt^\dagger \bt + \br^\dagger \br = \bm{1}$. 

\section{\label{sec:excluding}Excluding evanescent modes}

\begin{table}
\caption{\label{tab:times} CPU times in seconds when using WFM for
calculating $\bt$ and $\br$ at 20
different energies inside $E \in [-2~\mathrm{eV};2~\mathrm{eV}]$ 
for various two-probe systems. The numbers of atoms in the central region
(electrode unit cell) are indicated. 
The four right-most columns show the CPU times spent for 
computing the electrode bulk modes with \textsc{dgeev} and in this
work vs. solving the central region linear systems in \eq{eqn:linear}
and the system with two extra principal layers on each side.
}
\begin{ruledtabular}
\begin{tabular}{cc|cc|cc}
System & Atoms & \eq{eqn:linear} & \eq{eqn:linear}($l = 2$) 
& \textsc{dgeev} & This work \\
\hline
Fe--MgO--Fe              &      27(6) &  0.8 &  0.9 &  1.3 &  1.1\\
Al--C$\times$7--Al       &     74(18) &  0.4 &  0.6 &  3.6 &  1.6\\
Au--DTB--Au              &    102(27) &  8.1 & 13.5 & 91.0 & 28.2\\
Au--CNT(8,0)$\times$1--Au&    140(27) & 11.4 & 16.6 & 77.6 & 17.1\\
Au--CNT(8,0)$\times$5--Au&    268(27) & 45.3 & 50.3 & 83.6 & 17.8\\
CNT(8,0)--CNT(8,0)       &    192(64) &  7.0 & 11.9 &129.0 & 19.4\\
CNT(4,4)--CNT(8,0)       &256(64$|$64)&  7.2 & 12.4 &121.5 & 21.0\\
CNT(5,0)--CNT(10,0)      &300(40$|$80)& 24.7 & 31.5 &113.3 & 22.6\\
CNT(18,0)--CNT(18,0)     &   576(144) &172.2 &225.5 &1362.2&253.3\\
CNTFET (see \fig{fig:nt84-10}) &  1440(160) &259.8 &286.9 &4633.0&372.3\\
\end{tabular}
\end{ruledtabular}
\end{table}

The most time consuming task of the WFM method
is often to determine the electrode modes,
which requires solving a quadratic eigenvalue
problem.\cite{Ando1991}
As examples, see the profiling results listed in 
\tab{tab:times}, where we have used the method to
compute $\bt$ and $\br$ for a selection
of two-probe systems.
\footnote{
We should point out that the metallic
electrodes in the two-probe systems considered in
\tab{tab:times} can be fully described by much 
smaller unit cells than indicated
(often only a few atoms are needed) and therefore the time
spend on computing the bulk modes can be vastly reduced in these specific
cases. For a general method, however, which supports CNTs, nano
wires, etc. as electrodes, the timings are appropriate for showing the
overall trend in the computational costs.}
The CPU timings show that to determine the electrode modes
by employing the state-of-the-art {\sc lapack} 
eigensolver {\sc dgeev}
is, in general, much more expensive
than to solve the system of linear equations 
in \eq{eqn:linear}. We expect this trend to hold for larger 
systems as well. Therefore, in the attempt to model significantly larger
devices (thousands of atoms), it is of essential interest 
to reduce the numerical cost of the electrode modes calculation.
We argue that a computationally reasonable approach is to limit the number of 
electrode modes taken into account, e.g., by excluding
the least important evanescent modes. In this section,
a proper technique to do this in a rigorous and systematic fashion
is presented.

\subsection{\label{sec:decay}Decay of evanescent modes}

The procedure to determine the Bloch factors $\lambda_k$
and non-trivial modes $\bm{\phi}_k$ of an ideal electrode and subsequently
characterize these as right-going ($+$) or left-going ($-$) is
well described in the 
literature.\cite{Ando1991, Krstic2002, Khomyakov2005, Brocks2007}
We note that only the obtained
propagating modes with $|\lambda_k|=1$ are able to carry charge
deeply into the electrodes and thus enter the Landauer expression in
\eq{eqn:T}. The evanescent modes with $|\lambda_k| \neq 1$,
on the other hand, decay exponentially but can still contribute
to the current in a two-probe system, as the ``tails''
may reach across the central region boundaries.

\begin{figure}[tbp]
\begin{center}
\includegraphics*[width=8.3cm]{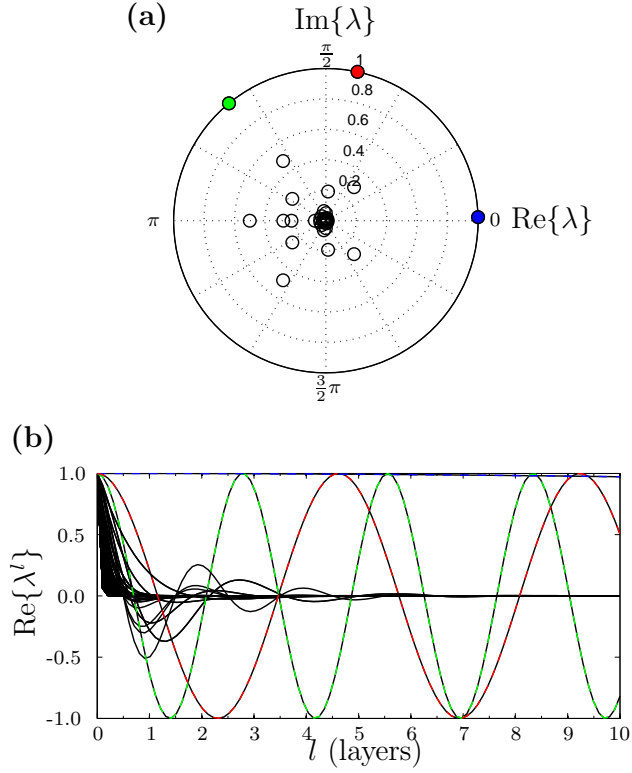}
\end{center}
\caption{(Color online)
(a) Positions of the Bloch factors $\lambda_k$ ($|\lambda_k| \le 1$)
obtained for a bulk Au(111) electrode with 27 atoms
per unit cell at $E=-1.5~\mathrm{eV}$. 
(b) Amplitudes of the corresponding normalized electrode modes $\bm{\phi}_k$ 
moving through 10 layers of the ideal bulk electrode.
A total of 243 modes are shown of
which 3 are propagating (colored/dashed) and the rest
are evanescent (circles/black).
}
\label{fig:waves}
\end{figure}

Consider a typical example of an electrode modes evaluation:
We look at a gold electrode with 27 atoms in the unit cell 
represented by 9 ($\mathrm{sp}^3\mathrm{d}^5$) orbitals for each Au-atom. 
Such a system results in 243 right-going and 243 left-going modes.
\fig{fig:waves}a shows the positions in the complex plane of
the Bloch factors corresponding to the right-going modes
(i.e.,  $|\lambda_k| \le 1$) for energy $E = -1.5~\mathrm{eV}$.
We see that there are exactly three propagating
modes, which have Bloch factors located on the unit circle.
The remaining modes are evanescent, of which many have Bloch factors
with small magnitude very close to the origin.

\fig{fig:waves}b illustrates how the 243 left-going
modes would propagate through 10 successive 
gold electrode unit cells. The figure shows that the amplitudes
of the three propagating modes are unchanged, while the
evanescent modes are decaying exponentially.
In particular, we note that the evanescent modes with Bloch factors
of small magnitude are very rapidly decaying and vanishes
in comparison to the propagating modes after only a few layers.
In the following, we will exploit this observation and
attempt to exclude such evanescent modes from the
WFM calculation altogether. Formally this can be accomplished if only the
electrode modes $\bm{\phi}_k$ with Bloch factors $\lambda_k$ satisfying
\begin{equation}\label{eqn:cond}
\lambda_{\min} \le |\lambda_k| \le \lambda_{\min}^{-1}, 
\end{equation}
are computed and subsequently taken into account, for 
a reasonable choice of $0 < \lambda_{\min} < 1$.
\eq{eqn:cond} is adopted as the key relation to
select a particular subset of the available electrode modes
(as recently suggested in Ref.~\onlinecite{Khomyakov2005}).

\subsection{\label{sec:extra}Extra electrode layers}

We will denote the mode, Bloch and self-energy matrices from which
the rapidly decaying evanescent modes are excluded with a tilde, 
i.e., as $\tilde{\bm{\Phi}}^{\pm}$, $\tilde{\bB}^{\pm}$
and $\tilde{\bm{\Sigma}}$. The mode matrices holding the
excluded modes are denoted by a math-ring accent
$\mathring{\bm{\Phi}}^{\pm}$, so that
\begin{equation}\label{eqn:split}
\bm{\Phi}^\pm = [\tilde{\bm{\Phi}}^\pm, \mathring{\bm{\Phi}}^\pm],
\end{equation}
is the assumed splitting of the full set.
All expressions to evaluate the Bloch and self-energy matrices are unchanged
as given in \sect{sec:formalism}
(now $(\tilde{\bm{\Phi}}^{\pm})^{-1}$ merely represents
the \textit{pseudo-inverses} of $\tilde{\bm{\Phi}}^{\pm}$).
However, since the column spaces of $\tilde{\bm{\Phi}}^{\pm}$ 
are not complete, there is no longer any guaranty that 
WFM can be performed so that
the resulting self-energy matrices and, in turn, the solution
$\bm{\psi}_C = [\bm{\psi}_{1}^\mathrm{T},
\dots, \bm{\psi}_{n}^\mathrm{T}]^\mathrm{T}$ of the
linear system in \eq{eqn:linear}, are correct.
In addition, it is clear that errors can occur
in the calculation of $\bt$ and
$\br$ from \eqs{eqn:t} and (\ref{eqn:r}) because
the boundary wave functions $\bm{\psi}_1$
and $\bm{\psi}_n$ might not be fully represented in
the reduced sets $\tilde{\bm{\Phi}}_R^{+}$ and $\tilde{\bm{\Phi}}_L^{-}$.

\begin{figure}[tbp]
\begin{center}
\includegraphics*[width=8.3cm]{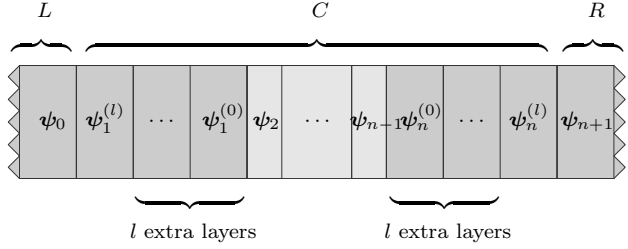}
\end{center}
\caption{(Color online)
Two-probe system in which the $C$ region boundaries 
are expanded by $l$ extra electrode layers.
}
\label{fig:wfm2}
\end{figure}

In order to diminish the errors introduced by excluding evanescent modes,
we propose to insert 
additional electrode layers in the central region,
see \fig{fig:wfm2}.
As illustrated in the previous section, 
this would quickly reduce the imprint of
the rapidly decaying evanescent modes in the boundary
layer wave functions
$\tilde{\bm{\psi}_1}$ and $\tilde{\bm{\psi}}_n$, which
means that the critical
components outside the column spaces
$\tilde{\bm{\Phi}}^\pm$ 
becomes negligible at an exponential rate in
terms of the number of additional layers.
We emphasize that the inserted layers may be ``fictitious'' in the sense that 
they can be accommodated by simple
block-Gaussian-eliminations prior to the solving of
\eq{eqn:linear} for the original system.

The above statements are confirmed by the following analysis. We expand the electrode wave functions  in the
corresponding complete set of bulk modes
\begin{equation}\label{eqn:expand}
\bm{\psi}_{i}^{\pm}
= \bm{\Phi}^\pm \ba_{i}^\pm
= [\tilde{\bm{\Phi}}^\pm, \mathring{\bm{\Phi}}^\pm]
\left[
\begin{array}{c}
\tilde{\ba}_{i}^\pm\\
\mathring{\ba}_{i}^\pm\\
\end{array}
\right],
\end{equation}
where $\ba_{i}^\pm = [
\tilde{\ba}_{i}^{\pm T},
\mathring{\ba}_{i}^{\pm T}]^\mathrm{T}$
are vectors that contain the expansion coefficients. In the particular case, where
$l$ extra electrode layers are inserted
and the border layers of the $C$ region are identical to
the connecting electrode layers, the electrode wavefunctions entering the matching 
 boundary  equations will be
\begin{equation}\label{eqn:bound1}
\bm{\psi}_{1}^{(l)-} = 
(\bB_L^-)^{-l} \bm{\psi}_{1}^{-} =
[\tilde{\bm{\Phi}}_L^-, \mathring{\bm{\Phi}}_L^-]
\left[
\begin{array}{c}
(\tilde{\bm{\Lambda}}_L^-)^{-l}\tilde{\ba}_{1}^-\\
(\mathring{\bm{\Lambda}}_L^-)^{-l}\mathring{\ba}_{1}^-\\
\end{array}
\right],
\end{equation}
and
\begin{equation}\label{eqn:bound2}
\bm{\psi}_{n}^{(l)+} = 
(\bB_R^\pm)^{l} \bm{\psi}_{n}^{+} =
[\tilde{\bm{\Phi}}_R^+, \mathring{\bm{\Phi}}_R^+]
\left[
\begin{array}{c}
(\tilde{\bm{\Lambda}}_R^+)^{l}\tilde{\ba}_{n}^+\\
(\mathring{\bm{\Lambda}}_R^+)^{l}\mathring{\ba}_{n}^+\\
\end{array}
\right],
\end{equation}
using the definition
$\bB^\pm = \bm{\Phi}^{\pm} 
\bm{\Lambda}^{\pm} (\bm{\Phi}^{\pm})^{-1}$. 
This shows that the critical components outside
the column spaces of $\tilde{\bm{\Phi}}_L^\pm$
and $\tilde{\bm{\Phi}}_R^\pm$ are given by
coefficients 
$(\mathring{\bm{\Lambda}}_L^-)^{-l}\mathring{\ba}_{1}^-$
and 
$(\mathring{\bm{\Lambda}}_R^+)^{l}\mathring{\ba}_{n}^+$, 
respectively. If this set only consists of the most rapidly decaying of
the evanescent modes according
to \eq{eqn:cond}, that is,
$|\lambda_k| > \lambda_{\min}^{-1}$ for the diagonal
elements of $\mathring{\bm{\Lambda}}_L^-$ and
$|\lambda_k| < \lambda_{\min}$ 
for the diagonal elements of $\mathring{\bm{\Lambda}}_R^+$,
where $\lambda_{\min}$ is less than $1$, these coefficients
always decrease as a function of $l$. 

We conclude that WFM with the reduced set of modes approaches the exact case  if additional electrode layers are inserted and  the solution $\tilde{\bm{\psi}}_C$ obtained from
\eq{eqn:linear} approaches the correct solution $\bm{\psi}_C$ accordingly.

\subsection{\label{sec:accuracy}Accuracy}

As pointed out above, the exclusion of some of the
evanescent modes from the mode matrices $\bPhi^\pm$
will introduce errors because the column spaces in
$\tbPhi^\pm$ are incomplete. In this section we will estimate how this will 
influence the accuracy of the calculated transmission
and reflection coefficients in terms of the parameter $\lambda_{\min}$ and the number
$l$ of extra electrode layers. 

Consider first the accuracy of the transmission matrix $\bt$
in the case of the extended two-probe system in \fig{fig:wfm2}.
For a specific
incoming mode $k$, we compare the correct result 
obtained with the complete set of modes (cf. \eq{eqn:t}),
\begin{equation}
\bt_{k} = 
\begin{bmatrix}
\tbt_{k}\\
\rbt_{k}\\
\end{bmatrix}
= [\tbPhi_R^+,\mathring{\bPhi}_R^+]^{-1}
\bm{\psi}_n^{(l)+},
\end{equation}
with the result obtained with the reduced mode matrix (denoted by a prime),
\begin{equation}
\bt_{k}^\prime=
\begin{bmatrix}
\tbt_{k}^\prime\\
\mathring{\bm{0}}^\prime\\
\end{bmatrix}
= [\tbPhi_R^+,\mathring{\bm{0}}]^{-1} \bm{\psi}_n^{(l)+},
\end{equation}
where 
$\mathring{\bm{0}}^\prime$ represents the zero vector of size 
$\mathring{m}_R$ and
$\mathring{\bm{0}}$ the  zero matrix of size  $m_R \times
\mathring{m}_R$.

The important coefficients in $\bt_k$ and  $\bt_k^\prime$ for 
transmission calculations are the ones representing the Bloch modes
which enters the Landauer-Büttiker formula in \eq{eqn:T}. Since
these are never excluded they will always be located
within the first $\tilde{m}_R$
elements, i.e., in $\tbt_k$ and  $\tbt_k^\prime$.
It then suffices to compare these parts of the transmission matrix
which we can do as follows.

From the properties of the pseudo-inverse we are able to write the
relation
\begin{equation}\label{eqn:pseudo}
(\tbPhi_R^+)^{-1}[\tbPhi_R^+,
\mathring{\bPhi}_R^+] = [\tilde{\bI},
(\tbPhi_R^+)^{-1}\mathring{\bPhi}_R^+],
\end{equation}
where $\tilde{\bI}$ is the identity matrix of order equal to the
number of included modes $\tilde{m}_R$.
Using the expression in \eq{eqn:bound2} it then follows that
\begin{equation}\label{eqn:ttilde}
\tbt_k=
(\tbLambda_R^+)^{l}\tba_{n}^+,
\end{equation}
and
\begin{equation}\label{eqn:tprime}
\tbt_k^\prime=
\tbt_k
+
(\tbPhi_R^+)^{-1}\mathring{\bPhi}_R^+
(\mathring{\bLambda}_R^+)^{l}\mathring{\ba}_{n}^+,
\end{equation}
where the $\tbt_k^\prime$ expression clearly corresponds to
the correct coefficients $\tbt_k$ plus an error term.

We have already established in the previous section that 
the $(\mathring{\bLambda}_R^+)^{l}\mathring{\ba}_{n}^+$
factor in the error term will
decrease as a function of $l$. We now show that the other term, $(\tbPhi_R^+)^{-1}\mathring{\bPhi}_R^+$
is independent of $l$, and consequently, that the error term in
\eq{eqn:tprime} must decrease as a function of $l$. To this end we look 
at the 2-norm of $(\tbPhi_R^+)^{-1}\mathring{\bPhi}_R^+$, which
satisfies
\begin{equation}\label{eqn:2norm}
|| (\tbPhi_R^+)^{-1}\mathring{\bPhi}_R^+ ||_2
\le
\mathring{m}_R^{\frac{1}{2}}
|| (\tbPhi_R^+)^{-1} ||_2,
\end{equation}
since $||\mathring{\bPhi}_R^+ ||_2 \le \mathring{m}_R^{\frac{1}{2}}$
when all evanescent modes are assumed to be normalized.
The norm $|| (\tbPhi_R^+)^{-1} ||_2$ can be 
readily evaluated and depends on the set of modes included
via the parameter $\lambda_{\min}$ but not on $l$. Thus, we conclude that the only  term of \eq{eqn:tprime} which depend on $l$ is $(\mathring{\bLambda}_R^+)^{l}\mathring{\ba}_{n}^+$, and the error is therefore decreasing as function of $l$.

Writing \eq{eqn:tprime} as $\tilde{\bt}_k^\prime=
\tilde{\bt}_k + \tilde{\bm{\epsilon}}_k$, where 
$\tilde{\bm{\epsilon}}_k$ holds the
errors on the coefficients of the $k$th
column, we further obtain that the total transmission $T^\prime$ 
can be expressed as
\begin{equation}\label{eqn:Tprime}
T^\prime = 
T + \sum_{kk^\prime} 
(\tilde{\mt}_{kk^\prime}^* \tilde{\epsilon}_{kk^\prime}
+\tilde{\epsilon}_{kk^\prime}^* \tilde{\mt}_{kk^\prime}
+|\tilde{\epsilon}_{kk^\prime}|^2)
\end{equation}
where $T$ is the exact result and the summation
is over the Bloch modes $k$ and $k^\prime$ in the
left and right electrode, respectively. 

For a first order estimate  of the error
term in \eq{eqn:Tprime} we consider 
the worst case approximation, where all diagonal elements of
$\mathring{\bm{\Lambda}}_R^+$ are equal to the
maximum range $\lambda_{\min}$
of \eq{eqn:cond}. This  makes all elements
$\tilde{\epsilon}_{kk^\prime}$ proportional to $\lambda_{\min}^{l}$.
and we arrive at the simple relation
\begin{equation}\label{eqn:estimate}
|T^\prime - T| \sim \lambda_{\min}^{l} + {\cal O}\big((\lambda_{\min}^{l})^2\big),
\end{equation}
which shows that the error 
decreases exponentially in terms of the
number of extra layers $l$.

For a higher order estimate of the error, we  directly monitor the error arising
on the boundary conditions,  
in terms of the coefficient vectors
$\tilde{\bm{b}}_{L,k} \equiv
(\tilde{\bm{\Phi}}_R^+)^{-1} (\bm{\psi}_1^{(l)+}
-\lambda_{L,k}^{+} \bm{\phi}_{L,k}^{+})$
and
$\tilde{\bm{b}}_{R,k} \equiv
(\tilde{\bm{\Phi}}_R^-)^{-1} \bm{\psi}_n^{(l)-}$, where 
$\bm{\psi}_1^{(l)+}$ and $\bm{\psi}_n^{(l)-}$ are given
by solving \eq{eqn:linear}. When the boundary conditions are exactly satisfied, we have
$|\tilde{\bm{b}}_{L,k}| = 0$ and $|\tilde{\bm{b}}_{R,k}| = 0$.
In the case where the boundary conditions are not exactly satisfied, 
$\tilde{\bb}_{R,k}$ represents the error on the left-going
components within the right boundary layer in the same way
that $\tilde{\bm{\epsilon}}_k$ represents the error on the 
right-going (transmitted) components. 
We would therefore expect the same order of magnitude
of $|\tilde{\bb}_{R,k}|$ and $|\tilde{\bm{\epsilon}}_k|$
in an actual calculation for a given mode $k$.
This suggests the following error estimate
from \eq{eqn:Tprime}, 
\begin{equation}\label{eqn:estimate2}
|T^\prime - T| \le 
\sum_{k}
(2|\tilde{\bt}_{k}| |\tilde{\bm{\epsilon}}_{k}|
+|\tilde{\bm{\epsilon}}_{k}|^2)
\sim
\sum_{k}
(2|\tilde{\bt}_{k}| |\tilde{\bm{b}}_{R,k}|
+|\tilde{\bm{b}}_{R,k}|^2),
\end{equation}
where all the vector norms
(e.g., $|\tilde{\bt}_k|^2 = \sum_{k^\prime} 
|\tilde{\mt}_{kk^\prime}|^2$)
are assumed to be taken over the elements
corresponding to Bloch bulk modes $k^\prime$ only.

Finally, we note without explicit derivation, that 
similar arguments
for the reflection matrix with columns $\tilde{\br}_k^\prime=
(\tilde{\bm{\Phi}}_L^-)^{-1}
(\bm{\psi}_1^{(l)-}
-\lambda_{L,k}^{+} \bm{\phi}_{L,k}^{+})$
and the total reflection coefficient $R^\prime$, 
results in the same accuracy expressions
for $|R^\prime - R|$ if we substitute
$\tilde{\bt}_{k} \rightarrow \tilde{\br}_{k}$ 
and $\tilde{\bm{b}}_{R,k} \rightarrow \tilde{\bm{b}}_{L,k}$  in \eqs{eqn:estimate}
and (\ref{eqn:estimate2}).

\subsection{\label{sec:example}Example}
\begin{figure}[tbp]
\begin{center}
\includegraphics*[width=12.0cm]{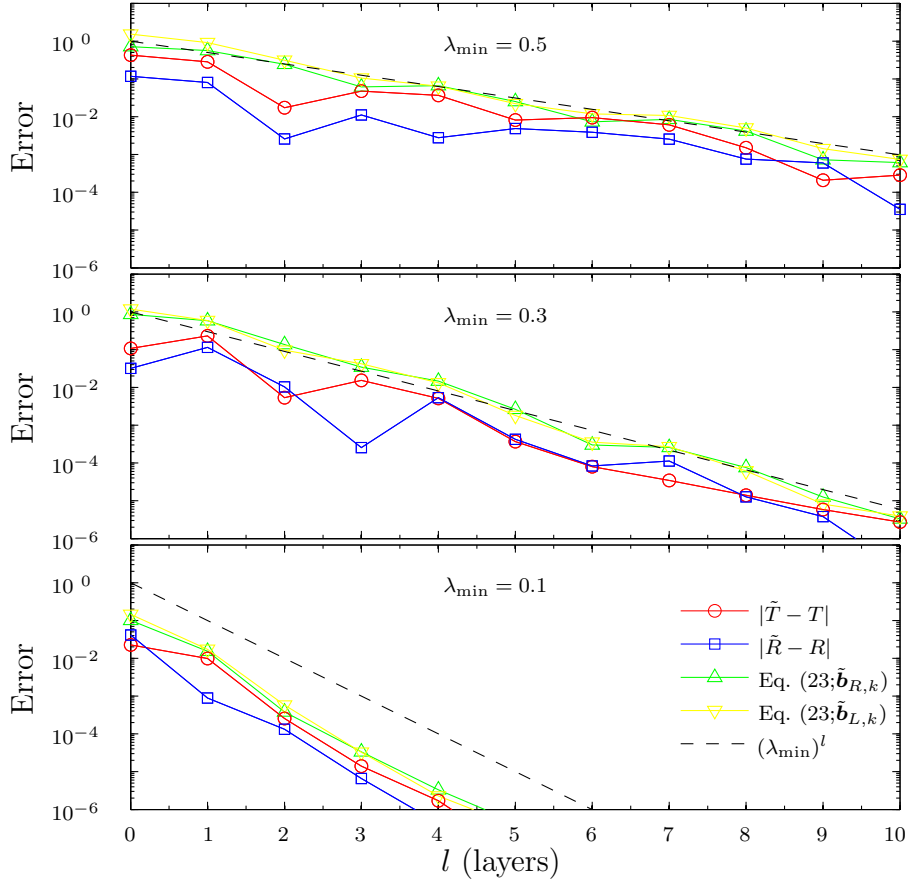}
\end{center}
\caption{(Color online)
Error (absolute) in the calculated total transmission (solid red lines) and reflection (solid blue lines) coefficients
$T^\prime$ and $R^\prime$ as a function of $l$. The panels show the cases of 
$\lambda_{\min}$ set to $0.5$, $0.3$ and $0.1$, which corresponds to
3, 14 and 31 Au bulk modes (out of 243, see \fig{fig:waves}) 
taken into account, respectively.
The dashed line indicates the 
first order  error estimate $\lambda_{\min}^l$. The yellow and green lines show error estimates obtained from \eq{eqn:estimate2}.
}
\label{fig:error}
\end{figure}

To end this section, we exemplify the previous discussion
quantitatively by looking at the Au(111) electrode described
earlier, and assuming a 128 atom (4 unit cells) device
of zigzag-(8,0) carbon nano tube (CNT) sandwiched between the gold electrodes, see the
configuration in \fig{fig:twoprobe}.
For energy $E=-1.5~\mathrm{eV}$,
we have calculated the deviation between
the total transmission obtained
when all bulk modes are
taken into account ($T$) and when some evanescent modes are
excluded ($T^\prime$) as specified with different settings of $\lambda_{\min}$.
Deviations are also determined for the corresponding
total reflection coefficients ($R$ and $R^\prime$).
\fig{fig:error} shows the results as a function of $l$, 
together with the estimate $\lambda_{\min}^{l}$ of \eq{eqn:estimate}
and the estimate of \eq{eqn:estimate2} both for the transmission and
reflection coefficients, 
where the higher order terms have been neglected, 

We observe that the absolute error in the obtained transmission
coefficients (red curves) and reflection coefficients (blue curves) 
are generally decreasing as a function of $l$,
following the same convergence rate as $\lambda_{\min}^{l}$ 
(dashed line). Looking closer at results for neighbor $l$ values, 
we see that the errors initially exhibit wave-like oscillations.
This is directly related to the wave form
of the evanescent modes that have been excluded
(see the propagation of the slowest decaying
black curves in \fig{fig:waves}(b)).
In other words, although the norm of the errors $|\tilde{\bm{\epsilon}}_k|$
are decreasing as a function of $l$, the specific error
$\tilde{\epsilon}_{kk^\prime}$ on a given 
(large) coefficient of $\tilde{t}_{kk^\prime}^\prime$
or
$\tilde{r}_{kk^\prime}^\prime$ may increase, which means that
the overall error term in \eq{eqn:Tprime} can go up.
Fortunately this is only a local phenomenon with the global
trend being rapidly decreasing errors.

Consider also the quality of the
simple accuracy estimate of $\lambda_{\min}^{l}$
and the estimates expressed by \eq{eqn:estimate2} for the
transmission coefficients (green curves) 
and reflection coefficients (yellow curves), respectively.
For relatively large $\lambda_{\min}$ all
estimates are very good. 
However, for smaller values of $\lambda_{\min}$,
only the latter two retain a high quality while
the $\lambda_{\min}^{l}$ estimate tends to be overly pessimistic.
It is important to remember that these estimates are by no means
strict conditions but in practice give very reasonable estimates of the accuracy.

We note in passing, that the results in the top panel of \fig{fig:error} 
corresponds to
using \textit{only} the propagating Bloch modes in the
transmission calculation. Still
we are able to compute
$T$ and $R$ to an absolute accuracy of three
digits by inserting $2\times 5$ extra
electrode layers in the two-probe system. 
This is quite remarkable and shows promise for large-scale
systems, e.g., with nano-wire electrodes, for which the 
total number of evanescent modes 
available becomes exceedingly great.

\section{\label{sec:numerical}Application}

In this section we will apply the developed method to a nano-device
consisting of a CNT stretched between to two metal electrodes
and controlled by three gates. 
The setup is inspired by 
Appenzeller \textit{et al.},\cite{Appenzeller2004} 
and we expect this particular arrangement
to be able to display so-called band-to-band (BTB) tunneling,
where one observes gate induced tunneling from the valence band 
into the conduction band of a semi-conducting CNT and vice versa.

\begin{figure}[tbp]
\begin{center}
\includegraphics*[width=\linewidth]{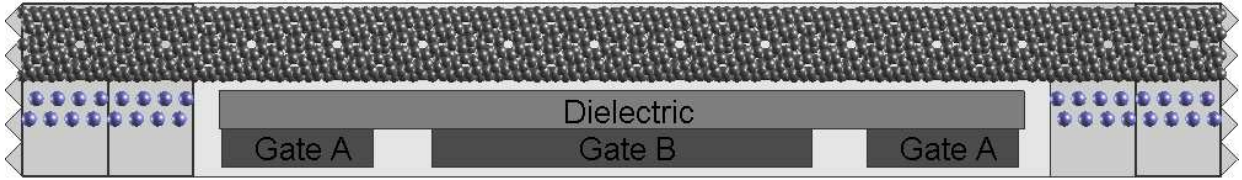}
\end{center}
\caption{(Color online)
Schematic illustration of a
carbon nanotube (8,4) band-to-band tunneling device.
The carbon nanotube is positioned on Li surfaces
next to an arrangement of three gates.
}
\label{fig:nt84-10}
\end{figure}

We show the configuration of the two-probe system in \fig{fig:nt84-10}.
The device configuration contains 10 principal layers
of a CNT(8,4), having 112 atoms in each layer. The diameter of the
tube and the thickness of the principal layer are 8.3~Å and 11.3~Å, respectively.
The electrodes consist of CNT(8,4) resting on
a thin surfaces of Li, where the lattice constant of the Li layers
is stretched to fit the layer thickness of the CNT.
The central region of the two-probe system comprises
a total of 1440 atoms. 
An arrangement of rectangular gates are positioned below 
the carbon nanotube as indicated on the figure.
In the plane of the illustration (length $\times$ height) the 
dimensions are as follows: Dielectric 108~Å $\times$ 5~Å; Gate-A 108~Å $\times$ 5~Å; 
Gate-B 20~Å $\times$ 5~Å. We set $\epsilon=4$ for the dielectric
constant of the dielectric in order to simulate SiO$_2$ or Al$_2$O$_3$
oxides. All the regions are centered with respect to the electrodes so
that the complete setup has mirror symmetry in the length
direction. In the direction perpendicular to the illustration the
configuration is assumed repeated every 19.5~Å as a super-cell. 

We have obtained the density matrix of the
BTB device by combining the NEGF formalism 
with a semi-empirical Extended Hückel model (EHT)
using the parameterization of Hoffmann.\cite{Hoffmann1963} From the
density matrix we calculate Mulliken populations on each atom, and
represent the total density of the system as a superposition of
Gaussian distributions on each atom properly weighted by the Mulliken
population. The width of the Gaussian is chosen to be consistent with
CNDO parameters.\cite{datta} The electrostatic interaction between the
charge distribution and the dielectrics and gates is subsequently calculated.
The Hartree-like term is then included in the Hamiltonian and the
combined set of equations are solved self-consistently. The 
resulting self-consistent EHT model is closely related to the work of
Ref.~\onlinecite{datta}, and a detailed description of the model
will be presented elsewhere.\cite{Stokbro2008}

In order to adjust the charge transfer between the CNT and
the Li electrodes we add the term $\delta \epsilon \mathrm{S}$ 
to the Li parameters. With an appropriate adjusted value of  $\delta
\epsilon$, the carbon nanotube becomes n-type doped. We adjust the
value such that the 
average charge transfer from Li to the
nanotube at self-consistency is $0.002$~e per carbon atom in the
electrode. The Fermi energy is then located at $-4.29~\mathrm{eV}$, which is
$0.07~\mathrm{eV}$ below the conduction band of the CNT(8,4).

In the following we fix $V_\mathrm{Gate-A}=-2.0$~eV and vary the
Gate-B potentials in the range 
$[-2~\mathrm{eV},4~\mathrm{eV}]$. Note that we report the gate potentials as an external  potential on
the electrons, and to translate the values into a gate potential  of unit Volts the values must be
divided with $-e$.

\begin{widetext}
\begin{figure}[tbp]
\begin{center}
\includegraphics*[width=0.8\linewidth]{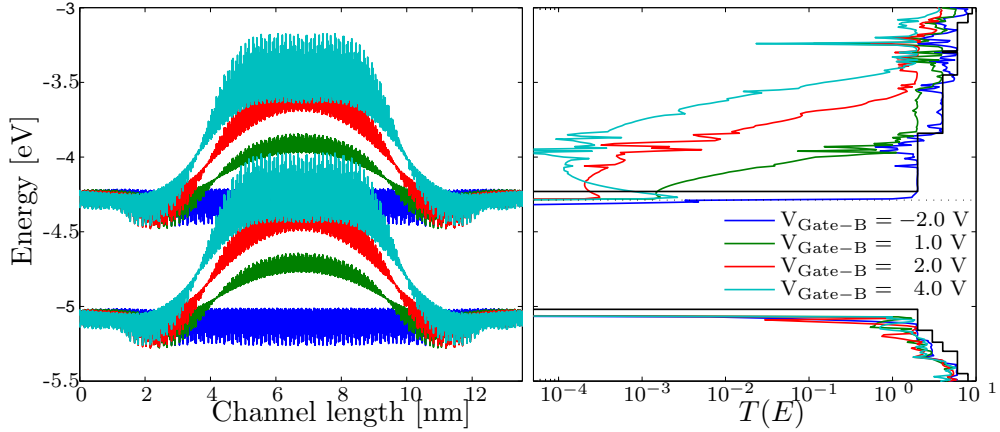}
\end{center}
\caption{(Color online)
Left panel: Representation of the electrostatic induced shift of the valence and conduction band edges along
the length of the device  for 
gate potentials $V_\mathrm{Gate-B} = $  $-2.0$~eV,  $1.0$~eV,  $2.0$~eV and
$4.0$~eV. Right panel: The corresponding transmission spectrum. The dotted line shows the position of the Fermi level, and
the solid line shows the transmission coefficient for an ideal CNT(8,4).
}
\label{fig:trans}
\end{figure}
\end{widetext}

In the left part of \fig{fig:trans} we present the total self-consistent
potential induced 
by the three gates on the carbon atoms 
in the CNT over the full extension of the device. For each configuration of the gate potentials the electrostatic
potential is shown twice, i.e., by two curves with the same color displaced relative to each
other with the  energy of the valence band and conduction band edge,
respectively. In this way the curves not only represent the
electrostatic potential of the device, but also the position of the
valence and conduction band edges.

Along with this,
in the right part of \fig{fig:trans}, we show the corresponding
transmission spectrum $T(E)$, for four gate potentials
$V_\mathrm{Gate-B}=-2.0$~eV, 1.0~eV, 2.0~eV, and 4.0~eV.
When $V_\mathrm{Gate-B}=-2.0$~eV the nanotube is largely 
unpertubed by the gate and the transmission coefficient is close to  
an ideal (8,4) CNT. We note that this is in agreement with ab initio 
calculations by  Nardelli et. al.,\cite{Nardelli2001} which found that a two 
terminal (5,5) CNT device in a similar contact geometry showed a 
nearly ideal conductance spectrum. In addition, the calculated band gap of the 
(8,4) nanotube is 0.81 eV, which is in good agreement with the 
value of 0.96 eV obtained from ab initio density-functional 
calculations in the generalized gradient approximation.\cite{Zhao2004} 

From \fig{fig:trans} we  see how the bands are shifted upwards 
by an increasing amount as the Gate-B potential is turned up.
To begin with, e.g., for $V_\mathrm{Gate-B}=1$~eV, 
this results in lower conduction since the conduction
band bends away from the Fermi level and the Fermi energy electrons need to tunnel
through the central region.  When the gate voltage is at $V_\mathrm{Gate-B}=2$~eV,
the valence band almost reaches the conduction band
in which case BTB tunneling becomes possible.
By increasing the gate voltage further, more bands become available for 
BTB tunneling and the effect is visible as a steady increase in the
calculated transmission $T(E)$ just above the Fermi level.

\begin{figure}[tbp]
\begin{center}
\includegraphics*[width=7cm]{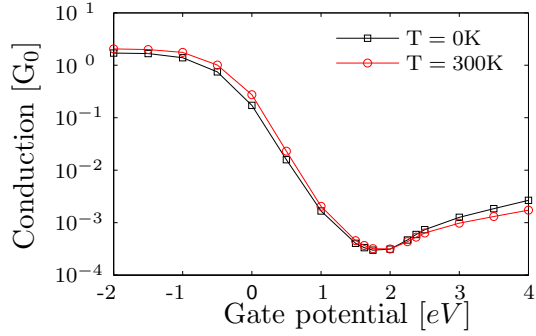}
\end{center}
\caption{(Color online)
Conduction 
in units of the conductance quantum $\mG_0$ as a function of the Gate-B potential.
In the calculations we use 
a dielectric constant of 4,
$V_\mathrm{Gate-A}=-2.0$~eV, and vary 
$V_\mathrm{Gate-B}$ from $-2.0$~eV to $4.0$~eV as indicated.
}
\label{fig:conduction}
\end{figure}

The results for the Fermi level
transmission $T(E_F)$ corresponding to the 
$\mT = 0~\mK$  unit conduction $\mG_0$,  are displayed
with the black curve in \fig{fig:conduction}. It shows an initial conductance
for $V_\mathrm{Gate-B}=-2.0$~V of the order of one,
a subsequent drop by four orders of magnitude around 
$V_\mathrm{Gate-B}=2.0$~V, and a final increase of one order of
magnitude towards $V_\mathrm{Gate-B}=4.0$~V. We also display the results
for the room temperature $\mT = 300~\mK$ conductance(red curve), which can be obtained
from 
\begin{equation}
\mG = \int \mathrm{d}E\ T(E)  \frac{e^{(E-E_F)/\mk_B \mT}}
{(1+e^{(E-E_F)/\mk_B \mT})^2}.
\end{equation}
The two conduction curves are similar, showing that the device is operating in
the tunneling regime rather than the thermal emission regime.

We next briefly comment on the comparison of the
simulation with the experiment of 
Appenzeller \textit{et al.}. \cite{Appenzeller2004}
In both cases the conduction curves have two
branches, which we denote Field Emission (FE) and Band to Band
Tunneling (BTB). Initially, the conduction decreases with applied gate
potential due to the formation of a barrier in the central region,
this is the FE regime. For 
larger biases the conduction increases again due to BTB tunneling, this is
the BTB regime. The experimental device display thermal
emission conduction and shows a corresponding  subthreshold slope, $S$, of $k_B T ln
(10)/e\approx 60$ mV/dec in the FE
regime.  The theoretical device, on the other hand, display tunneling conduction
and has $S \approx 500$ mV/dec in the FE regime. In the BTB regime,
the theoretical device has   $S \approx 2000$ mV/dec, while the
experimental device show $S \approx 40$ mV/dec. 

The very different
behavior is due to the short channel length of the theoretical
device. The central barrier has a length of $\approx 5$ nm, and at
this  length the electron can still tunnel through the barrier. We see
that the short channel length not only affects the subthreshold slope
of the FE regime, but also strongly influence the BTB regime.
Work are in progress for a parallel implementation of the methodology,
which will make it feasible to simulate larger systems, and thereby investigate the transition
from the tunnelling to the thermal emission regime.

\begin{figure}[tbp]
\begin{center}
\includegraphics*[width=8cm]{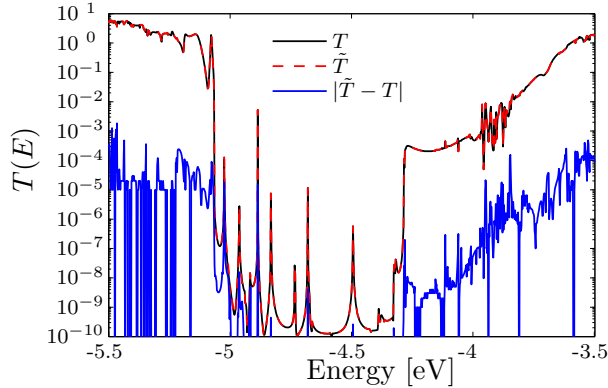}
\end{center}
\caption{(Color online)
Transmission coefficients $T$ and $\tilde{T}$ calculated with the
standard WFM method (black solid)
and the method of this work (red dashed), respectively, and the
difference $|\tilde{T}-T|$ (blue line) as a function of energy $E$ in the 
$V_\mathrm{Gate-B}=2$~V case.
}
\label{fig:comp}
\end{figure}

All the above results have been calculated with the modified WFM method using parameters
$\lambda_{\min} = 0.1$ and $l=1$. Thus, the results presents a
non-trivial application of the new method. To verify the
transmission results in \fig{fig:trans} we present a comparison
with the standard WFM method in \fig{fig:comp}. The figure shows that the 
transmissions curves are identical to about three significant digits.
The CPU time required for calculating a complete transmission
spectrum for \fig{fig:trans} is ($\sim 3$ hours), while the
corresponding calculation presented in \fig{fig:comp} with the 
standard WFM method took ($\sim 35$ hours).
Thus, the overall time saving achieved with the new method was therefore more than an order 
of magnitude. The results in \tab{tab:times} indicate that similar
timesavings can be expected for other systems with non-trivial electrodes.

\section{\label{sec:summary}Summary}
 
We have developed an efficient approach for calculating
quantum transport in nano-scale systems based on the
WFM scheme originally proposed by Ando in reference 
[\onlinecite{Ando1991}].
In the standard implementation of the WFM method
for two-probe systems, all bulk modes of the electrodes
are required in order to represent the transmitted and
reflected waves in a complete basis. 
By extending the central region of the two-probe system
with extra electrode principal layers, we are able to
exclude the vast majority of the evanescent bulk modes from the
calculation altogether. Our final
algorithm is therefore highly efficient,
and most importantly, errors and accuracy can be closely monitored.

We have applied the developed WFM algorithm to a CNTFET
in order to study the mechanisms of band-to-band
tunneling. The setup was inspired by reference
[\onlinecite{Appenzeller2004}], and the calculation display 
features also observed in the experiment, however, due to the short
channel length the theoretical device operates in the tunneling regime, while
the experimental device operates in the thermal emission regime.

By measuring the CPU-times for calculating transmission spectra
of the CNTFET two-probe system and comparing to the cost of the standard WFM
method we have observed a speed-up by more than a factor of 10. We see
similar speed-up for other non-trivial systems.
We therefore believe that this
is an ideal method to be used with ab-initio transport
schemes for large-scale simulations.

\begin{acknowledgments}\label{sec:acknowledgements}
This work was supported by the 
Da\-nish Council for Strategic Research (NABIIT)
under grant number 2106--04--0017, 
``Parallel Algorithms for
Computational Nano--Science''.
\end{acknowledgments}



\end{document}